\documentclass[aps,prl,fleqn,twocolumn,superscriptaddress,showpacs]{revtex4}

\newcommand{\bed}{\[}
\newcommand{\eed}{\]}
\newcommand{\beq}{\begin{equation}}
\newcommand{\eeq}{\end{equation}}
\newcommand{\beqa}{\begin{eqnarray}}
\newcommand{\eeqa}{\end{eqnarray}}
\newcommand{\ket} [1] {\vert #1 \rangle}
\newcommand{\bra} [1] {\langle #1 \vert}

\usepackage{graphicx}
\usepackage[mathscr]{eucal}
\usepackage{amsmath}
\usepackage{amssymb}

\usepackage{epstopdf}
\usepackage{epsfig}
\begin{document}

\title{Multipartite Asymmetric Quantum Cloning}

\author{S. Iblisdir}
\affiliation{GAP-Optique, University of Geneva,
20 rue de l'Ecole-de-M\'edecine, CH-1211, Switzerland}

\author{ A. Ac\'{\i}n}
\affiliation{ICFO-Institut de Ci\`encies Fot\`oniques, Jordi Girona 29,
08034 Barcelona, Spain}

\author{N. Gisin}
\affiliation{GAP-Optique, University of Geneva,
20 rue de l'Ecole-de-M\'edecine, CH-1211, Switzerland}

\author{J. Fiur\'{a}\v{s}ek}
\affiliation{QUIC, Ecole Polytechnique, CP 165,
Universit\'{e} Libre de Bruxelles, 1050 Brussels, Belgium }
\affiliation{Department of Optics, Palack\'{y} University,
17. listopadu 50, 77200 Olomouc, Czech Republic}

\author{R. Filip}
\affiliation{Department of Optics, Palack\'{y} University,
17. listopadu 50, 77200 Olomouc, Czech Republic}

\author{N.J. Cerf\,}
\affiliation{QUIC, Ecole Polytechnique, CP 165,
Universit\'{e} Libre de Bruxelles, 1050 Brussels, Belgium }


\begin{abstract}
We investigate the optimal distribution of quantum information over
multipartite systems in asymmetric settings. We introduce 
cloning transformations that take $N$ identical replicas of a pure
state in any dimension as input, and yield a collection of clones
with non-identical fidelities. As an example, if the clones are partitioned
into a set of $M_A$ clones with fidelity $F^A$ and another set of $M_B$ clones with fidelity $F^B$, the trade-off between these fidelities is
analyzed, and particular cases of optimal $N \to M_A+M_B$ cloning
machines are exhibited. We also present an optimal 
$1 \to 1+1+1$ cloning machine, which is the first known 
example of a tripartite fully asymmetric cloner. Finally, it is shown
how these cloning machines can be optically realized.
\end{abstract}

\pacs{03.67.-a, 03.65.-w}

\maketitle

A most intriguing feature of quantum mechanics is the
impossibility to clone perfectly the state of a quantum system
\cite{woot82}, generally referred to as the ``no-cloning theorem''.
This impossibility, which is deeply related to the impossibility
of superluminal communication \cite{diek82},
marks a fundamental difference between quantum
and classical information, and has far-reaching implications. It
puts strong limitations on the way we can protect quantum
information \cite{preskill},  and is related to another ``no-go''
result of quantum mechanics, namely the impossibility to acquire 
full information about the state of a quantum system from
a finite number of copies \cite{brus98}. Although it may first sound
negative, this no-cloning theorem can be turned as an advantage,
as beautifully demonstrated by quantum cryptography.
Two parties, willing to communicate privately, 
can exploit this impossibility to securely encode classical information
in quantum systems. Any intervention of an eavesdropper 
causes a disturbance, making it visible (see for example \cite{gisi02}
and references therein).
\par

One then understands the importance of going beyond the no-cloning
theorem and to formulate quantitatively the impossibility of
cloning, that is, to address the problem: ``{\it What is the best possible
operation to produce approximate copies 
of a given quantum state}?''. The simplest instance of this problem is
the duplication of a qubit, as was considered in \cite{buze96}
and demonstrated experimentally in \cite{experiments}. 
Many  generalizations and variants have followed, such as
the issue of producing $M$ clones from
$N$ replicas of a qubit \cite{gisi97}, the generalization to {\em qudits}
($d$-level quantum systems) \cite{wern98,cerf-aps,keyl98} and continuous-variable systems \cite{cerf00}, the cloning of non-identical 
states \cite{fiur02}, or non-universal cloning
\cite{dari01}. Much attention has also been
devoted to asymmetric  $1\rightarrow 1+1$  cloning machines,
which produce two clones with different fidelities $F^A$ and $F^B$
\cite{cerf-aps,cerf00:asym}. This allows, for instance, to investigate the
trade-off between the information obtained by an eavesdropper and the
disturbance of the receiver's state when cloning is used as an
attack on quantum key distribution.
\par

The present work introduces the concept of asymmetric cloning 
in a multipartite situation. A general asymmetric  $N \to M_A+ \ldots+ M_P$
cloning machine can be defined as a transformation that produces
$M=\sum_j M_j$ clones out of $N$ identical replicas of an unknown pure
state, such that the clones are partitioned into $P$ parties, 
consisting of $\{M_A,\ldots,M_P\}$ clones having the fidelities $\{F^A,\ldots,F^P\}$. We want to characterize the
trade-off between these fidelities. Such machines
have at least two applications. First, some $N \to M_A+M_B$
cloning machines have been proven to be a useful tool when
investigating the security of some quantum key distribution
schemes \cite{acin04}. Second, in the case where $M_A=N$
and $M_B \to \infty$, this allows us to study the
trade-off between the gain of knowledge about the state of
a quantum system and the disturbance undergone by this system.
More generally, these machines can be used to elucidate how
quantum information can be distributed unequally over many parties.
\par

In this paper, we illustrate this general scenario of multipartite asymmetric cloning with two examples, namely  (i) the optimal $1\to 1+n$ qubit cloning machines, and (ii) the optimal $1 \to 1+1+1$ {\em qudit} cloning machines. In both cases, an optical scheme realizing the optimal cloning based on parametric down-conversion is exhibited in the case of qubits. A general analysis of multipartite asymmetric cloning going beyond these examples will be reported in \cite{ibli04:clon,fiur04:clon}.
\par

\paragraph{Preliminaries.}
The cloning  operation is described by a trace-preserving
completely positive (CP) map $\mathcal{S}$ \cite{wern98}. We
characterize the quality of the clones with single-clone
fidelities. For the first set of clones, which we label A, the
single-clone fidelity reads
\[
\label{eq:deflocalfid}
F^A(\mathcal{S}) =
\min_{\psi \in \mathscr{H}} \, \min_{1 \leq k \leq
M_A} \bra{\psi} \mathrm{Tr}_{k}^{\prime}[\mathcal{S}(\ket{\psi^{\otimes N}}
\bra{\psi^{\otimes N}})] \ket{\psi}.
\]
In this expression, $\mathrm{Tr}_{k}^\prime$
denotes the partial trace over all clones but the $k$-th in the
first set. A similar expression holds for the single-clone
fidelity $F^B(\mathcal{S})$ of the second set of clones B, etc.
In the case where there are only two sets of clones, A and B, 
the problem of finding the optimal  $N \to M_A+M_B$ asymmetric cloner 
simply boils down to maximizing $F^B(\mathcal{S})$ for a given value of
$F^A(\mathcal{S})$. This calculation, as well
as the proof of optimality, will be presented in details 
in \cite{ibli04:clon,fiur04:clon}. Here, we only sketch 
the idea behind this optimization. 
Let $\ket{\psi}$ denote the state of each original we
want to clone. As demonstrated in \cite{keyl98}, the effect of an
optimal $N \to M$ symmetric cloning machine for {\em qudits} is to
produce clones whose individual state reads $\rho=\eta
\ket{\psi}\bra{\psi}+(1-\eta) \openone / d,$ for some
constant $\eta$ called the ``shrinking'' factor. This isotropy property
results from the fact that no state is preferred by an optimal universal
cloning machine. Therefore, the quality of the cloning process is
completely characterized by $\eta$. This also
applies to \emph{asymmetric} universal cloning. 
If we require all the $M_A$ clones to be of quality $\eta^A$, 
and all the $M_B$ clones to be of quality $\eta^B$, then the problem of 
optimal asymmetric cloning reduces to finding some tight relation between
$\eta^A$ and $\eta^B$. This can be done by exploiting the same techniques
as those used in \cite{keyl98} for symmetric cloners, that is, the Stinespring
representation of CP maps, the U(d)-covariance and permutation invariance.
As an example of this technique, we treat below the $1\to 1+n$
cloning of qubits. Alternatively, using the isomorphism between CP maps and 
positive semidefinite operators, the optimal asymmetric cloning machines 
can be found by analyzing the eigenstates and eigenvalues of certain operators. This second approach is illustrated below 
on the example of the $1 \rightarrow 1+1+1$
cloning of {\em qudits}.

\paragraph{Optimal $1\to 1+n$ universal cloning of qubits.} 

According to the Clebsch-Gordan series $j_1 \otimes j_2 \approx |j_1-j_2| \oplus \ldots \oplus (j_1+j_2)$,
the Hilbert space associated to $n+1$ qubits decomposes into irreducible subspaces as $\mathcal{H}^+_{n+1} \oplus \mathcal{H}^+_{n-1}$. Let $S_{n+1}$ and $S_{n-1}$ denote the corresponding projectors. We have found that optimal  $1 \to 1+n$ machines are of the form
\bed
T: \rho \to (\alpha^* S_{n+1}+\beta^* S_{n-1})(\rho \otimes \openone^{\otimes n})
(\alpha S_{n+1}+\beta S_{n-1}).
\eed
This form naturally generalises symmetric cloners found in \cite{wern98}. The corresponding fidelities can be conveniently written as 
\beq\label{eq:onetoneven}
F^A  = 1-\frac{2}{3}y^2, \qquad
F^B  = \frac{1}{2}+\frac{1}{3n}(y^2+\sqrt{n(n+2)}xy),
\eeq
where $x^2+y^2=1$. (The relation between $(\alpha,\beta)$ and $(x,y)$ is irrelevant for our discussion.)
Note that these expressions are only valid for $n>1$;  for $n=1$, the optimal cloning machines are those discussed in \cite{cerf-aps,cerf00:asym}. One readily checks that imposing $F^A=1$ implies $F^B=1/2$. This is consistent with the idea that in order to prepare a perfect clone, one has to take it from the input, which therefore cannot interact with any system \cite{jozsa02}. Then, no quantum information is available to prepare the $n$ extra clones, and the best one can do is to prepare $n$ random states with fidelity $1/2$. If, instead, one requires the $n$ clones to have the fidelity of an optimal symmetric $1 \to n$ cloning machine, namely $F^B=(2n+1)/(3n)$, one finds $F^A=(2n+1)/[3(n+1)]$. Interestingly, this fidelity is larger than $1/2$ for $n>1$: not all quantum information need be used to produce the $n$ optimal clones, and some quantum information remains to prepare a non-trivial $n+1$-th clone.
\par

This $1\to 1+n$ cloner is also interesting in the limit of large $n$
in the context of the connection between cloning and state
estimation \cite{gisi97,brus98}. For universal symmetric cloning, 
it is known that there is a one-to-one
correspondence between the $n \to \infty$ cloning machines and
state estimation devices \cite{brus98:esti}. Such a relation still
holds in the asymmetric case. Following the lines of
\cite{brus98:esti}, one finds that, in the limit $n \to \infty$,
asymmetric $1 \to 1+n$ cloning machines interpolate between a
(trivial) machine, leaving the quantum system unchanged, and a
fully measuring device, estimating destructively the input state. 
Indeed, for $n\to\infty$, Eqs.~(\ref{eq:onetoneven}) become
\begin{equation}
F^A = 1-\frac{2}{3} y^2, \qquad F^{\textrm{meas}} =
\frac{1}{2}+\frac{1}{3}y\sqrt{1-y^2},
\label{state-estim}
\end{equation}
where only the range $0 \leq y \leq 1/\sqrt{2}$ is relevant.
If $F^A=1$, then one finds $F^{\textrm{meas}} = 1/2 $, which
means that no information can be gained if the input
state is unperturbed.
The maximum value of $F^{\textrm{meas}}$ is 2/3, which is consistent with
\cite{mass95}. In that case, of course, $F^A=2/3$. In between these
two cases, Eqs.~(\ref{state-estim}) express the optimal trade-off between the
knowledge gained on a system and the disturbance effected by the measurement.
This trade-off had previously been studied in the form 
of an inequality \cite{bana01}.
Our machine provides a concrete means to realize measurements 
saturating this inequality.

\paragraph{Optimal $1\to 1+1+1$ universal cloning of qudits.}
We now consider fully asymmetric tripartite  universal
cloning machines in dimension $d$, which produce clones $A$, $B$, 
and $C$ with respective fidelities $F^A$, $F^B$,
and $F^C$. The optimal cloners should be such that 
for a given pair of fidelities
(say, $F^A$ and $F^B$), the fidelity of the third clone ($F^C$) is
maximal. This is equivalent to maximizing a convex mixture of the
fidelities, $F=a F^{A}+b F^{B} + cF^{C}$, for fixed $a,b,c\geq 0$
with $a+b+c=1$. The tripartite asymmetry is then controlled 
by the ratios $a/b$ and $a/c$.
\par

The optimal cloning transformation can be determined by exploring the
isomorphism between trace-preserving CP maps $\mathcal{S}$
and positive  semidefinite operators $S$ on the tensor product
of input and output Hilbert spaces.
The mean fidelity $F^A$ of the first clone averaged over all
input states $|\psi\rangle$ can be expressed as $F^A=\mathrm{Tr}[SL_A]$,
where $L_A=\int_{\psi} \psi_{\mathrm{in}}^T \otimes \psi_A \otimes \openone_B
\otimes \openone_C \, d \psi,$ $T$ stands for transposition and
$\psi=|\psi\rangle\langle \psi|.$ The fidelities $F^B$ and $F^C$ can be
expressed similarly, and one has to maximize $F=\mathrm{Tr}[SL]$, where
$L=a L_A+b L_B+c L_C$. The fidelity $F$ is upper bounded by the maximum eigenvalue $\lambda_{\mathrm{max}}$ of  $L$, $F\leq d \lambda_{\mathrm{max}}$
\cite{Fiurasek01}. This bound happens to be tight here, and is saturated by the
optimal cloner. The operator $S$
describing this cloner is therefore proportional to the projector 
onto the subspace spanned by the eigenstates of $L$ with
eigenvalue $\lambda_{\mathrm{max}}$.
\par

A unitary implementation of this tripartite asymmetric
cloner requires two ancillas, $E$ and $F$,  so that
any pure input state $|\psi\rangle$ transforms according to
\begin{eqnarray}
|\psi\rangle &\rightarrow &
\mathcal{C}\left[ \alpha
|\psi\rangle_{A}(|\Phi^{+}\rangle_{BE}|\Phi^{+}\rangle_{CF}
+|\Phi^{+}\rangle_{BF}|\Phi^{+}\rangle_{CE}) \right. \nonumber \\
&&+\beta|\psi\rangle_{B}(|\Phi^{+}\rangle_{AE}|\Phi^{+}\rangle_{CF}
+|\Phi^{+}\rangle_{AF}|\Phi^{+}\rangle_{CE}) \nonumber \\
&&\left.+\gamma |\psi\rangle_{C}(|\Phi^{+}\rangle_{AE}|\Phi^{+}\rangle_{BF}
+|\Phi^{+}\rangle_{AF}|\Phi^{+}\rangle_{BE}) \right]. \nonumber \\
\label{onetothree}
\end{eqnarray}
Here, $\mathcal{C}=\sqrt{d/[2(d+1)]}$ is a normalization constant, $|\Phi^{+}\rangle=d^{-1/2}\sum_{j=1}^d|j\rangle|j\rangle$
is a maximally-entangled state of two {\em qudits},
and $\alpha,\beta,\gamma\geq 0$ obey
\begin{equation}
\alpha^2+\beta^2+\gamma^2+\frac{2}{d}(\alpha\beta+\alpha\gamma+\beta\gamma)=1.
\label{alfabetagammanorm}
\end{equation}
The transformation (\ref{onetothree}) is universal, that is, the
single-clone fidelities do not depend on the input state and can be expressed
in terms of the coefficients $\alpha$, $\beta$, and $\gamma$ as
\begin{equation}
\begin{array}{l}
F^{A}=1-\frac{d-1}{d}\left[\beta^2+\gamma^2+\frac{2\beta\gamma}{d+1}\right],
\\[2mm]
F^{B}=1-\frac{d-1}{d}\left[\alpha^2+\gamma^2+\frac{2\alpha\gamma}{d+1}\right],\\[2mm]
F^{C}=1-\frac{d-1}{d}\left[\alpha^2+\beta^2+\frac{2\alpha\beta}{d+1}\right].
\end{array}
\end{equation}
In the special case where $\gamma=0$, the expressions of $F^A$ and $F^B$
exactly coincide with those for the $1\to 1+1$ asymmetric cloners
found in \cite{cerf-aps,cerf00:asym}. This actually confirms the optimality
of these cloners, which was only conjectured previously for $d>2$.
Note that the third clone has a fidelity $F^C$ exceeding $1/d$ here, which
is again related to the information left in the anti-clones.
\par

\paragraph{Optical implementations.}
We now focus on optical implementations in which the mechanism
responsible for cloning is stimulated emission \cite{SWZ}. The
optical scheme consists of a parametric down-conversion (PDC)
process stimulated by $N$ identical photons injected in the signal mode.
Taking pulsed type-II frequency degenerated PDC whose Hamiltonian
reads $H=\gamma(a_{VS}^\dagger a_{HI}^\dagger-
    a_{HS}^\dagger a_{VI}^\dagger) + \textrm{ h.\,c. }$,
and assuming that the photon to be cloned is in the state 
$(\alpha a_{VS}^\dagger+\beta a_{HS}^\dagger)\ket{\textrm{vac}}$ 
with $\ket{\textrm{vac}}$ denoting the vacuum state, 
one obtains for the state after the crystal
\begin{equation}\label{stim}
    \ket{\psi_s}=e^{-iHt}(\alpha
a_{VS}^\dagger+\beta a_{HS}^\dagger)^N\ket{\textrm{vac}}.
\end{equation}
Postselecting the events when $M$ photons are detected in the signal
mode, one recovers the optimal fidelities of the symmetric $N\to M$
cloning machine for qubits \cite{SWZ}. Note that when
$M$ photons are observed in the signal mode, $N$ of them come from
the initial state while $M-N$ were produced in the crystal, which
means that there are $M-N$ photons in the idler mode too (these are 
the anti-clones). This scheme can be modified in order to implement 
the asymmetric $1\to 1+1$ machine \cite{Filip}. After a successful 
$1\to 2$ symmetric cloning (that is, when one photon pair is generated),
the two clones are split at a first beam splitter, and one of them 
is combined with the anti-clone at a second beam splitter 
of transmittance $T$ in order to break the symmetry between the clones.
Long but straightforward algebra shows that the fidelities at the modes $A$
and $B$, depending on $T$, are those of the $1\to 1+1$ 
machine \cite{cerf-aps,cerf00:asym}.

\begin{figure}[h]
\begin{center}
\epsfig{figure=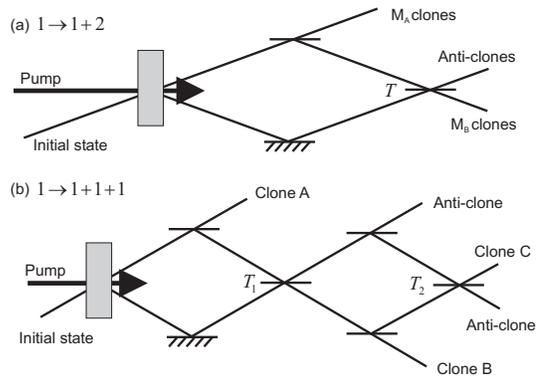,width=70mm}
\caption{Optical implementation of (a) the $1\to 1+2$ 
and (b) the $1\to 1+1+1$ optimal asymmetric cloning machines.}
\label{condimpl}
\end{center}
\end{figure}

\begin{figure}[h]
\begin{center}
\epsfig{figure=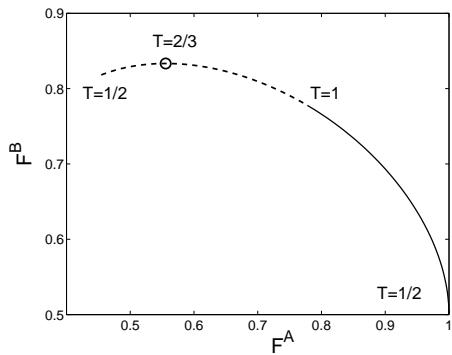,width=60mm}
\caption{Clone fidelities for the optical implementation of the
 $1\to 1+2$ optimal machine. The solid (resp.  dashed) line
 represents the case when $M_A=1$ and $M_B=2$ (resp.  $M_A=2$
 and $M_B=1$) photons are postselected.}\label{figexpfid112}
\end{center}
\end{figure}

This modified scheme can be extended to implement the
$1\to 1+2$ asymmetric cloning machine introduced above, 
as shown in Fig.~\ref{condimpl}(a). (The case $n>2$ will be
discussed in subsequent papers.) The idea is again to postselect
the events when there are 5 photons (3 in the signal mode, and 2
in the idler mode) and then to combine some of signal photons 
with idler photons at a beam splitter of transmittance $T$.
By postselecting the cases where $M_A=1$ and $M_B=2$, 
one gets the fidelities
\beq
\label{expfid1121} 
F^A=\frac{4T^2-4T+7}{12T^2-12T+9}; \quad
F^B=\frac{8T^2-4T+3}{12T^2-12T+9} 
\eeq 
as shown in Fig.~\ref{figexpfid112} (solid line, for 
$1/2\leq T\leq 1$). The optimal symmetric machine is
recovered for $T=1$. If $T$ decreases, the quality
of the first clone in mode $A$ increases, while the quality 
of the two clones in mode $B$ decreases. When $T=1/2$, 
all the information in mode $B$ is lost, and a perfect copy 
of the initial state is obtained in mode $A$.
The trade-off between the two fidelities exactly follows the
optimal $1\to 1+2$ machine predicted by Eq.~(\ref{eq:onetoneven}).
However, it is not possible to recover the entire curve because Eq.~(\ref{expfid1121}) implies $F^A \geq F^B$. To get the remaining
$1 \to 1+2$ cloning machines for which $F^A \leq F^B$, one
uses the same setup but postselects the events with
$M_A=2$ and $M_B=1$. We then find the fidelities \beq\label{expfid1122}
F^A=\frac{7T^2-4T+4}{9T^2-12T+12}; \quad
F^B=\frac{3T^2-4T+8}{9T^2-12T+12} . \eeq
The corresponding curve is also shown in Fig.~\ref{figexpfid112} (dashed line).
For example, when $T=2/3$ the two photons in mode $A$ have a fidelity $5/6$,
as in the case of the symmetric $1\to 2$ cloner,
while the photon in mode $B$ has fidelity $5/9$, slightly larger than
1/2 as discussed above.

This optical scheme can also be adapted to realize the optimal 
$1\to 1+1+1$ cloning transformation, see Fig.~\ref{condimpl}(b).
Here, the output of a symmetric $1\to 3$ machine
is made asymmetric by combining some of the clones with
anti-clones at two beam splitters, with transmittance $T_1$ and
$T_2$. After patient calculation, one can check
that the obtained fidelities $F^A\geq F^B\geq F^C$,
depending on $T_1$ and $T_2$, are
optimal. The three fidelities are equal when $T_1=T_2=1$.
Other interesting limiting cases are $(F^A,F^B,F^C)=(1,1/2,1/2)$
when $T_1=1/2$, while taking $T_2=1$ gives Eqs.~(\ref{expfid1121})
for the $1\to 1+2$ case. This construction can
easily be generalized further. 
\par

In summary, we have introduced the concept of multipartite asymmetric
quantum cloning machines, and have illustrated it with two examples. These devices have several remarkable properties. One of them is that if one wants to produce $n$ clones from an input with a fidelity which is as high as possible, some quantum information still remains to produce a nontrivial $n+1$-th clone. Multipartite cloning machines provide a new tool for analyzing the security of multipartite
quantum cryptography, and the tradeoff between disturbance and information gain. We have presented feasible optical realizations of our examples of optimal multipartite cloners. We have then seen that the noise that characterizes these cloning machines is related to the  unavoidable spontaneous emission that necessarily accompanies stimulated emission. One is tempted to conjecture that all the $N\to M_1+...+M_P$ cloning machines for qubits are only limited by spontaneous emission, and can therefore be implemented by splitting and then recombining the clones and anti-clones produced by stimulated emission using a series of beam splitters, in a similar way as in Fig. \ref{condimpl}. The corresponding fidelities would define the optimal distribution of a qubit among several parties. 
\par

Financial support by the EU projects SECOQC, RESQ and CHIC is acknowledged.
NJC and JF also acknowledge support from the Communaut\'e
Fran\c{c}aise de Belgique under grant ARC 00/05-251, and from the IUAP
programme of the Belgian government under grant V-18. 
JF and RF also acknowledge support from the
grant LN00A015 of the Czech Ministry of Education,
and AA acknowledges support from the Spanish MCYT 
under "Ram\'on y Cajal" grant.

\end{document}